\documentclass[prc,aps,superscriptaddress,showpacs,nofootinbib,twocolumn]{revtex4}
\usepackage{amssymb}
\usepackage[dvips]{graphicx}
\usepackage[english]{babel}
\usepackage{indentfirst}
\usepackage{amsxtra}
\usepackage{amsmath}
\usepackage{supertabular}
\usepackage{multirow}
\usepackage[mathcal]{eucal}
\usepackage[usenames]{color}
\usepackage{ulem}

\begin{document}
\title {\bf Tensor parameters in Skyrme and Gogny effective interactions: 
Trends from a ground--state--focused study}

\author{Marcella Grasso}
\affiliation{Institut de Physique Nucl\'eaire, IN2P3-CNRS, Universit\'e Paris-Sud, 
F-91406 Orsay Cedex, France}

\author{Marta Anguiano}
\affiliation{Departamento de F\'{\i}sica
At\'omica, Molecular y Nuclear, Universidad de Granada, 
E-18071 Granada, Spain}

\begin{abstract} 
Recent ground--state--focused studies of the tensor effects in the mean--field framework 
are our starting point. On the basis of phenomenological arguments, we indicate regions 
for acceptable values of the parameters that are associated with the tensor effective 
forces within both the Skyrme and the Gogny models. We identify acceptable signs and values 
of the parameters by making an adjustment on the neutron $1f$ spin--orbit splitting for 
the nuclei $^{40}$Ca, $^{48}$Ca and $^{56}$Ni. The first nucleus is not used to adjust the tensor parameters 
because it is spin--saturated, but is employed to tune the spin--orbit strength. One of the main conclusions of this work is 
that some existing Skyrme parametrizations containing the tensor force should not be employed because the wrong 
sign of the tensor parameters does not lead to the correct behavior (by comparing with the experimental results). 
This study also allows us to better constrain the tensor parameters in the Gogny case, where much less work is 
published and boundaries and signs for the parameters have not been analyzed so far.  

\end{abstract} 

\vskip 0.5cm \pacs {21.60.Jz, 21.10.-k,21.10.Pc} \maketitle 
%

\section{Introduction}

Tensor effects in mean--field--based theories with effective interactions have been extensively studied by several 
groups in the last decade. 

An early attempt to adjust the tensor parameters was performed in the 70s by Stancu et al. \cite{stancu77} on top of 
the Skyrme parametrization SIII \cite{beiner} and provided indicative boundaries for the values of the tensor 
parameters, that constitute the so--called Stancu--Brink--Flocard (SBF) triangle. Recently, more modern and 
upgraded parametrizations were proposed, some of them going beyond the limits fixed by the SBF triangle. 
Tensor parameters were introduced in Ref. \cite{brown2006}, based on the Skyrme parametrization Skx \cite{brown1998}. 
Brink and Stancu reconsidered their previous adjustment and provided new tensor parameters by including this time 
also single--particle observables of exotic nuclei in the adjustment procedure  \cite{stancu2007}. 
Col\`o et al. introduced tensor parameters \cite{colo} on top of the Skyrme interaction SLy5 \cite{chabanat}. 
Lesinski et al. adjusted 36 parametrizations by performing a global fit of all the Skyrme parameters including 
the tensor contribution \cite{lesinski}. New tensor parameters were proposed also in 
Refs. \cite{zalewski2008,zalewski2009}, where a special emphasis has been put on the importance of 
using single--particle properties in the adjustment of the parameters. 

Much less work has been published for finite--range interactions. In the interactions introduced by Nakada, a 
finite--range (Yukawa form) tensor term is included \cite{nakada}. On the other side, the tensor contribution is 
systematically disregarded in the currently used Gogny interactions \cite{gogny}, like D1S \cite{berger} and 
D1M \cite{goriely}. In the pioneering study of Onishi and Negele, a Gaussian tensor term was added to a Gaussian 
central term and its effects were analyzed \cite{onishi}. The first parametrization in the framework of the full 
Gogny interaction was introduced only very recently by Otsuka et al. \cite{otsuka}, by including a Gaussian 
tensor--isospin term and by refitting all the other Gogny parameters together with the tensor parameter. 
Also Co' et al. introduced a finite--range tensor--isospin term \cite{co}, but the tensor contribution was added on 
top of the existing D1S and D1M Gogny interactions. More recently, a pure Gaussian tensor term was included together 
with a tensor--isospin term of the same type to allow the authors to adjust in a separate and independent way 
the neutron--proton and the like--nucleon tensor contributions in the Gogny framework, as is commonly done in the Skyrme 
case \cite{anguiano}. It has been shown that the neutron--proton 
and the like--particle contributions are indeed proportional and have the same sign if only a tensor--isospin term is included. 
 
In this work, we start from some existing parametrizations with the objective of putting global 
phenomenological boundaries for the acceptable values of the parameters governing the neutron--proton and the 
like--particle contributions of the Skyrme and Gogny tensor forces. For the Skyrme case, the 
existing parametrizations of Col\`o et al. \cite{colo} (that we call SLy5$_{\rm T}$) and Brink et al. 
\cite{stancu2007} (SIII$_{\rm T}$), as well as four parametrizations taken from the work of Lesinski et al. 
\cite{lesinski} will be considered. As an illustration, we have chosen 
T$_{11}$, T$_{14}$, T$_{41}$ and T$_{44}$ because each of these parametrizations explores one of 
the possible signs of the two tensor parameters. For the Gogny case, there is only one study where 
the two tensor contributions have been studied independently \cite{anguiano}. Our starting point for the 
Gogny case will be the interaction D1ST2a introduced in Ref. \cite{anguiano}.  

We follow the same idea as in Refs. \cite{zalewski2008,zalewski2009} and analyze the neutron $1f_{7/2}-1f_{5/2}$ spin--orbit 
splittings in the three nuclei $^{40}$Ca, $^{48}$Ca and $^{56}$Ni. First, we check the $1f$ spin--orbit splitting 
in the spin--saturated nucleus $^{40}$Ca and  modify the spin--orbit parameter to better reproduce this quantity. 
We then check how well the neutron $1f$ spin--orbit splitting is reproduced in the nucleus $^{48}$Ca. In this 
case, only the like--particle contribution enters into play because this nucleus is spin--saturated in protons. 
We mention that this spin--orbit splitting has already been used in Ref. \cite{anguiano} to adjust the like--particle 
part of the Gogny tensor term. In this second step, the parameter governing the like--particle contribution is 
modified. Finally, we check the neutron--proton contribution of the tensor term in the nucleus $^{56}$Ni where both 
contributions are active and we adjust the corresponding parameter (the like--particle part, already fixed in the previous step, is kept fixed). This procedure 
follows what has been done in Ref. \cite{zalewski2009} where, in the context of the Energy Density Functional (EDF), 
the isoscalar tensor coupling constant $C^J_0$ has been adjusted on the $1f$ spin--orbit splitting of the nucleus 
$^{56}$Ni and the isovector tensor coupling constant $C^J_1$ has been adjusted on the same spin--orbit splitting 
in the nuleus  $^{48}$Ca. It has to be mentioned that, in Refs. \cite{zalewski2008,zalewski2009}, 
the single--particle energies are evaluated for even nuclei by using the theoretical mass of such nuclei and 
the masses of the neighboring odd nuclei. This allows the authors to perform a more meaningful comparison 
with the corresponding experimental single--particle energies that are extracted in the same way from the 
experimental masses. To do this, a correct treatment of odd nuclei has to be done by breaking the 
time--reversal symmetry. 
To compare with the experimental values we treat in this work only even nuclei, we keep the spherical symmetry and we  
use the single--particle Hartree--Fock (HF) 
energies. 

The present article is organized as follows. In Sec. II the meaning of the tensor parameters in the Skyrme 
and Gogny cases is briefly recalled. In Sec. III the $1f$ spin--orbit splittings are analyzed in the three 
nuclei $^{40}$Ca, $^{48}$Ca, and $^{56}$Ni and the tensor effects related to each parameter are studied. 
The spin--orbit paramater (tensor) parameters are modified to better reproduce the experimental values in 
$^{40}$Ca ($^{48}$Ca and $^{56}$Ca). Regions for acceptable values of the parameters are identified. 
Conclusions are reported in Sec. IV.

\section{Parameters}

In the following, we assume for simplicity spherical symmetry and perform HF calculations by neglecting pairing 
correlations. The correlations associated to the particle--vibration coupling are also neglected. The objective 
is not to perform accurate adjustments, but to provide some trends and to exclude, on the basis of phenomenological arguments, 
regions of parameter values that lead to an incorrect description of the tensor effects. 

\subsection{Skyrme case}

By considering the zero--range tensor term already included in the original work of Skyrme \cite{skyrme}, 
the so--called $J^2$ terms appear in the mean--field Hamiltonian density, where $J$ is the spin--orbit density. The 
two parameters governing these terms are often called $\alpha$ and $\beta$. They contain both a central--exchange 
contribution ($\alpha_{\rm C}$ and $\beta_{\rm C}$) and a tensor contribution ($\alpha_{\rm T}$ and $\beta_{\rm T}$): 
 \begin{eqnarray}
\nonumber
\alpha&=&\alpha_c+ \alpha_T, \\
\nonumber
\beta&=&\beta_c+\beta_T.
\end{eqnarray}
The parameters $\alpha_{\rm C}$ and $\beta_{\rm C}$ are expressed in terms of the Skyrme parameters of the 
velocity--dependent terms: 
\begin{eqnarray}
\nonumber
\alpha_{\rm C}&=&\frac{1}{8}\left ( t_1\, - \, t_2 \right )\,  - \, \frac{1}{8} 
\left ( t_1x_1 \, + \, t_2x_2 \right ), \\
\nonumber
\beta_{\rm C} &=&-\frac{1}{8} \left ( t_1x_1 \, + \, t_2x_2 \right ).
\end{eqnarray}
The parameters $\alpha_{\rm T}$ and $\beta_{\rm T}$ are written as:
\begin{eqnarray}
\nonumber
\alpha_{\rm T}&=&\frac{5}{12} \, U, \\
\nonumber
\beta_{\rm T}&=&\frac{5}{24} \left ( T+U \right ) ,
\end{eqnarray}
where $T$ and $U$ are the strengths of the Skyrme zero--range tensor force in even and odd states of relative motion, 
respectively \cite{skyrme};  $\alpha_{\rm T}$ describes the like--particle and $\beta_{\rm T}$ the neutron--proton tensor contributions.  The Hamiltonian density that is obtained within the variational HF scheme by starting 
from the Skyrme--Hamiltonian represents the so--called Skyrme energy functional, that is a functional of the local 
density. The coupling constants of this density functional are expressed in terms of the Skyrme parameters. 
On the other side, within an EDF perspective,  by writing down a functional of the local density in its most general 
form, the tensor contribution is governed by the so--called isoscalar $C^J_0$ and isovector $C^J_1$ coupling 
constants, that are related to the above mentioned parameters $\alpha$ and $\beta$ as follows: 
\begin{eqnarray}
\nonumber
\alpha&=&C^J_0 \, + \, C^J_1, \\
\nonumber
\beta&=&C^J_0 \, - \, C^J_1.
\end{eqnarray}

\subsection{Gogny case}

We consider a Gaussian tensor term like in Ref. \cite{anguiano}:  
\begin{equation}
V(r_1,r_2)= \left ( V_{{\rm T}1}+V_{{\rm T}2} P^{\tau}_{12} \right ) S_{12} {\rm exp}[-(r_1-r_2)^2/\mu^2_{\rm T}], 
\end{equation}
where $P^{\tau}_{12}$ is the isospin exchange operator and $S_{12}$ the usual tensor operator. 
The range $\mu_{\rm T}$ of this force has been chosen in Ref. \cite{anguiano} equal to the longest range of the 
D1S Gogny interaction, that is $1.2$ fm. The parameters $V_{{\rm T}1}$ and $V_{{\rm T}2}$ represent the strengths of 
the tensor force. In particular, $V_{{\rm T}1} + V_{{\rm T}2}$ is the strength of the force acting in like-nucleon 
pairs whereas $V_{{\rm T}2}$ represents the strength for the neutron--proton contribution. 

\section{Neutron  $1f_{7/2}-1f_{5/2}$  spin-orbit splitting}

Let us consider first the fully spin-saturated nucleus $^{40}$Ca.  The $1f$ spin--orbit splitting is not dependent on 
the tensor part of the force for this nucleus. This guarantees that the spin--orbit parameter can be 
tuned in a clean and proper way because it  represents the only contribution to the spin--orbit potential. In the upper panels of 
Fig. 1 we show the neutron $1f_{7/2}-1f_{5/2}$ spin--orbit splitting for the three nuclei $^{40}$Ca (a), $^{48}$Ca (b), 
and $^{56}$Ni (c) obtained with the chosen Skyrme forces and the  Gogny interaction of Ref. \cite{anguiano}. 
The experimental values are taken from Ref. \cite{schwierz}. 
It has to be mentioned that the neutron $f_{5/2}$ state is strongly fragmented in the nucleus $^{40}$Ca. For our adjustment, we have used the cetroid energy that is equal to 6.8 MeV and that is evaluated in Ref.  \cite{schwierz} by taking 24 states between 4.9 and 9.1 MeV. This centroid value is coherent with that reported in Ref. \cite{uozumi}. 

We observe in panel (a) that the spin--orbit splitting is systematically larger than the experimental value with all 
the used effective interactions for the nucleus $^{40}$Ca. 
We thus reduce in all cases the spin--orbit parameter of the force to obtain a value closer to the experimental 
splitting (d). The reduced spin--orbit parameters $W$ are reported in Table I. 
For simplicity, we employ along this work the same names for the forces, even if the parameters are 
refitted. 
\begin{table} 
\begin{tabular}{cccccccc}
\hline
            & SLy5$_{\rm T}$ & SIII$_{\rm T}$ & T$_{41}$ & T$_{11}$ & T$_{14}$ & T$_{44}$ & D1ST2a \\
\hline 
\hline
Old $W$     & 126.00         & 120.00         & 138.15   & 103.74   & 128.51   & 161.37   &  130.00 \\  
New $W$     &  101.00         & 95.50          & 103.00    & 100.00    & 101.00    & 105.00    &  103.00   \\
\hline
\end{tabular}
\caption {Upper line: standard spin--orbit parameters $W$ associated to each force. Lower line: reduced 
spin--orbit parameters $W$ obtained to reproduce the neutron $f$ spin--orbit splitting in the nucleus $^{40}$Ca. 
The reported values are in units of MeV fm$^5$.}
\end{table}

By using the reduced spin--orbit parameters, the neutron $f$ spin--orbit splittings are evaluated also for the 
other two nuclei and the corresponding results are reported in the panels (e) and (f) for the 
nuclei $^{48}$Ca and $^{56}$Ni, respectively. 

\begin{widetext}

\begin{figure}[htb]
\begin{center}
\includegraphics[width=12cm]{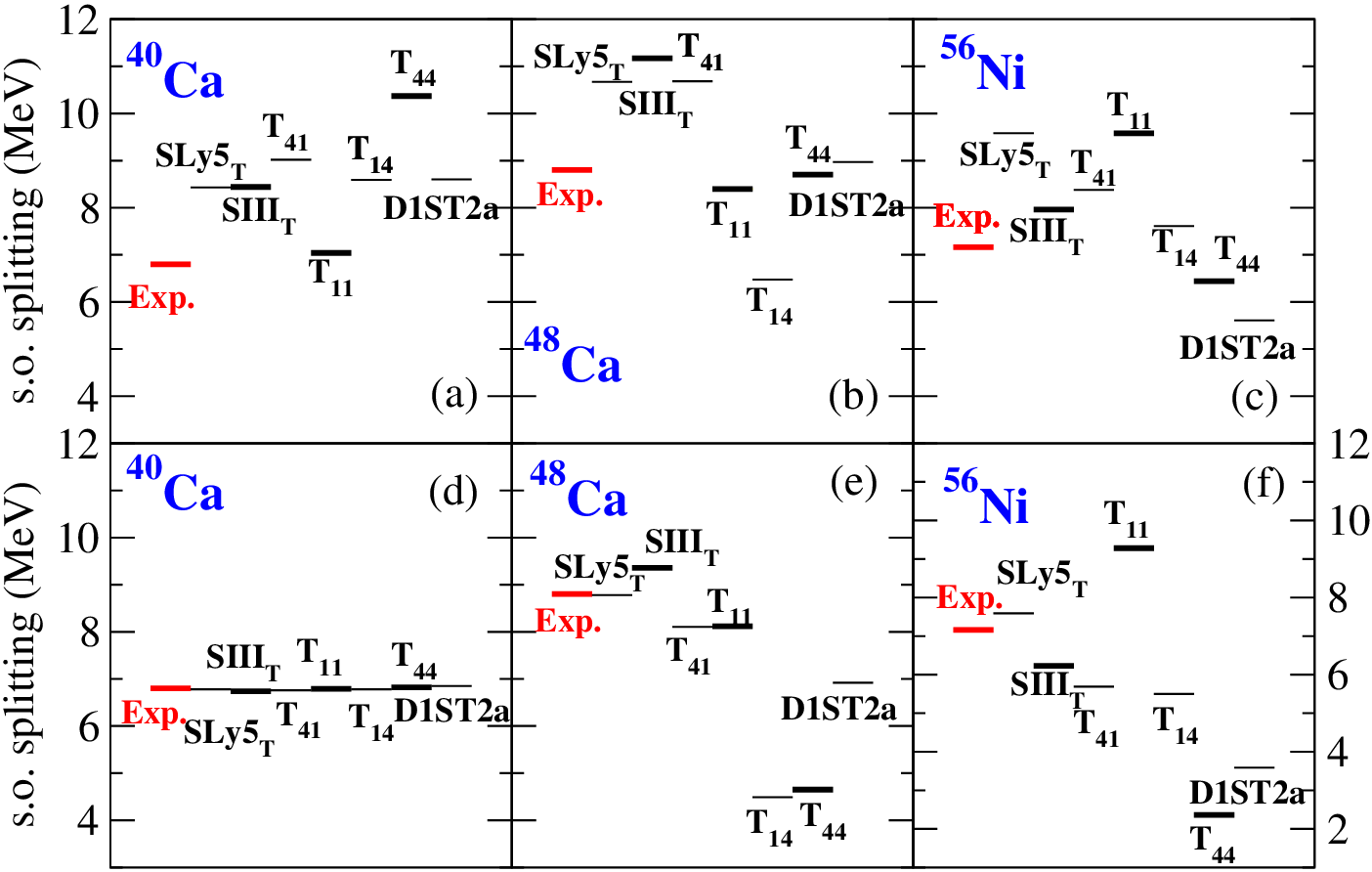}
\end{center}
\caption{ (Color online) Upper panels: neutron $f$ spin--orbit splittings calculated with several Skyrme and 
Gogny forces for the nuclei $^{40}$Ca (a), $^{48}$Ca (b), and $^{56}$Ni (c), respectively. The experimental 
values are also shown. (d), (e) and (f): same as in the upper panels, but calculated with the reduced spin--orbit 
parameters.}
\end{figure}

\end{widetext}

Let us first analyze the case of the nucleus $^{48}$Ca. This nucleus is spin--saturated in protons whereas it is not 
saturated for the neutrons. The tensor contribution to the Hamiltonian density and to the spin--orbit potential is 
due only to a like--particle neutron--neutron effect. There is thus a neutron--neutron tensor contribution to 
the $f$ spin--orbit splitting. We observe in panel (b) that the spin--orbit splitting 
(evaluated with the standard spin--orbit parameter) is in some cases larger and in other cases smaller than 
the experimental value. In the case of the Gogny interaction the spin--orbit splitting
is practically the same as the experimental value because the interaction D1ST2a was fitted in order 
to reproduce that value. When the reduced spin--orbit strength is used (e), the values are almost in all cases smaller 
than the experimental splitting, with the exception of SLy5$_{\rm T}$ and SIII$_{\rm T}$.  
The tensor force should now be used to better tune the theoretical splitting. 

To better analyze the effect of the tensor correlations, we plot in Fig. 2 the neutron $1f$ splitting 
obtained for the nucleus $^{48}$Ca by employing the reduced strength for the spin--orbit part (adjusted for the 
nucleus $^{40}$Ca) and by comparing the results obtained with (b) and without (a) the like--particle tensor 
contribution. In the latter case, the parameters $\alpha_{\rm T}$ for the Skyrme force and $V_{{\rm T}1}+V_{{\rm T}2}$ for 
the Gogny force are put equal to zero. The parameters $\beta_{\rm T}$ and $V_{{\rm T}2}$ may take any arbitrary 
value because their effect is in any case negligible. We keep for these parameters the same values associated to 
each original force.  When the tensor effects are quenched, the spin--orbit splitting is in all cases much lower 
than the experimental value (a). We observe that the action of the tensor force goes in the correct direction 
(towards the experimental value) for the interactions SLy5$_{\rm T}$, SIII$_{\rm T}$, T$_{41}$, T$_{11}$, and D1ST2a 
and in the opposite direction for $T_{14}$ and $T_{44}$. The first five cases correspond to negative values of 
$\alpha$ and $V_{{\rm T}1}+V_{{\rm T}2}$,  whereas in the last two Skyrme parametrizations $\alpha$ is positive. The 
effect for the case T$_{44}$ is very weak because the corresponding value of $\alpha_{T}$ is small. This first 
analysis indicates that the acceptable values of $\alpha$ and $V_{{\rm T}1}+V_{{\rm T}2}$ are those that lead to an 
increasing spin--orbit splitting for the nucleus $^{48}$Ca, that is, both $\alpha$ and $V_{{\rm T}1}+V_{{\rm T}2}$ 
negative. We thus keep only the interactions SLy5$_{\rm T}$, SIII$_{\rm T}$, T$_{41}$, T$_{11}$, and D1ST2a. For 
these cases, we modify the parameters $\alpha_{\rm T}$ (for Skyrme)  and $V_{{\rm T}1}+V_{{\rm T}2}$ (for Gogny) to 
better reproduce the experimental value of $8.8$ MeV. 
The results are shown in Fig. 3 and the new parameters are listed in Table II. 
\begin{figure}[htb]
\begin{center}
\includegraphics[width=8cm]{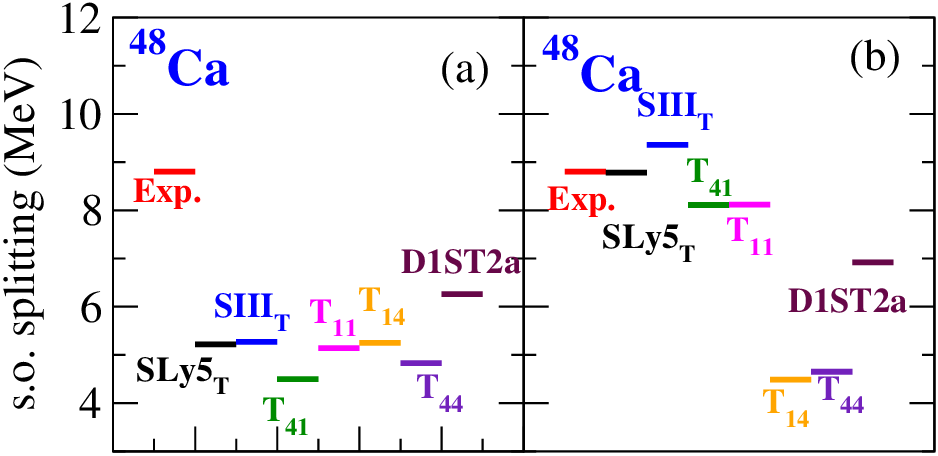}
\end{center}
\caption{ (Color online) (a): Spin--orbit splitting calculated by putting equal to zero the parameters 
$\alpha_{\rm T}$ and $V_{{\rm T}1} + V_{{\rm T}2}$. (b): Same as in panel (e) of Fig. 1.}
\end{figure}
\begin{figure}[htb]
\begin{center}
\includegraphics[width=8cm]{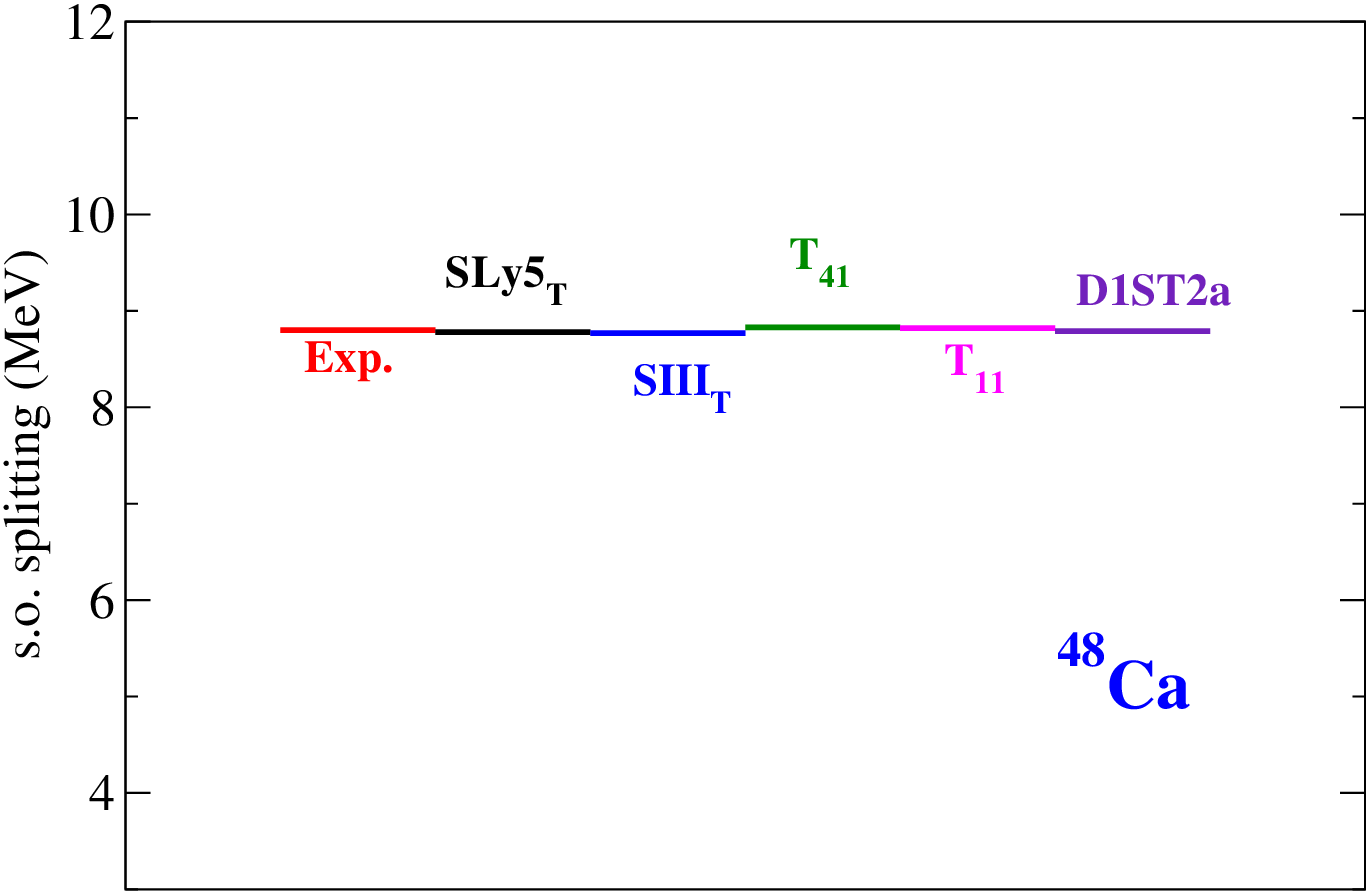}
\end{center}
\caption{ (Color online) Spin--orbit splitting calculated by refitting the parameters $\alpha_{\rm T}$ and 
$V_{{\rm T}1} + V_{{\rm T}2}$ to reproduce the experimental value.}
\end{figure}
\begin {table} 
\begin{tabular}{ccccc}
\hline
 SLy5$_{\rm T}$ & SIII$_{\rm T}$ & T$_{41}$ & T$_{11}$ & D1ST2a \\
\hline
\hline
-170.00         & -155.00        & -215.00  & -175.00  & -75.00 \\
\hline 
\end{tabular}
\caption {Tensor parameters $\alpha_{\rm T}$ (first four columns) and $V_{{\rm T}1}+V_{{\rm T}2}$ (last column) 
adjusted to reproduce the $1f$ splitting in the nucleus $^{48}$Ca. The reported values are in units of MeV fm$^5$.}
\end{table}
Let us now look at the nucleus $^{56}$Ni. From Fig. 1, we can make the same comments done for $^{48}$Ca. Some of the 
calculated $1f$ splittings are larger and others smaller than the experimental value in panel (c). When the reduced 
spin--orbit parameter is employed, almost all the values become smaller than the experimental value, with the 
exception of the SLy5$_{\rm T}$ and $T_{11}$ cases. By using now the reduced spin--orbit parameter and the refitted values for 
$\alpha_{\rm T}$ and $V_{{\rm T}1}+V_{{\rm T}2}$, we evaluate again the $f$ spin--orbit splittings for this nucleus 
with the interactions that have been selected in the previous step. The results are shown in Fig. 4 for 
$\beta_{\rm T}$ and $V_{{\rm T} 2}$ equal to zero (a) and different from zero (b). When the neutron--proton tensor 
effects are quenched (a) the splittings are in all cases larger than the experimental value. We observe in panel (b) that for the 
Skyrme cases  SLy5$_{\rm T}$, SIII$_{\rm T}$ and T$_{41}$, and for the Gogny case D1ST2a the tensor effects are correctly 
predicted leading to a reduction of the splitting, towards the experimental value. For the T$_{11}$ case, the 
splitting increases when the neutron--proton tensor contribution is taken into account. In the first four cases, 
the parameters $\beta$ and  $V_{{\rm T}2}$ are positive whereas in the T$_{11}$ case the parameter $\beta$ is 
negative. We can thus exclude the interaction  T$_{11}$. 

The parameters $\beta_{\rm T}$ and $V_{{\rm T}2}$ can now be adjusted to better reproduce the splitting in the 
nucleus $^{56}$Ni for the remaining three Skyrme cases and for the Gogny case. The refitted spin--orbit, 
$\alpha_{\rm T}$ and $V_{{\rm T}1}+V_{{\rm T}2}$ parameters are used.  The results are shown in Fig. 5 and the 
refitted parameters are listed in Table III. Also the old parameters are shown in Table III for comparison. 
\begin {table} 
\begin{tabular}{ccccc}
\hline
                                                   & SLy5$_{\rm T}$ & SIII$_{\rm T}$ & T$_{41}$ & D1ST2a \\
\hline 
\hline
Old $W$                                             & 126.00        & 120.00         & 138.15   &  130.00 \\  
New $W$                                             & 101.00         & 95.50          & 103.00    & 103.00 \\
\hline
Old $\alpha_{\rm T}$ or $V_{{\rm T}1}+V_{{\rm T}2}$ & -170.00       & -180.00        & -180.65  &  -20.00 \\  
New $\alpha_{\rm T}$ or $V_{{\rm T}1}+V_{{\rm T}2}$ & -170.00       & -155.00        &-215.00   & -75.00  \\
\hline
Old $\beta_{\rm T}$ or $V_{{\rm T}2}$               & 100.00        & 120.00         & 94.04    &  115.00 \\  
New $\beta_{\rm T}$ or $V_{{\rm T}2}$               & 122.00        & 50.00          & 55.00    &   60.00 \\
\hline
\end{tabular}
\caption {Old and refitted values for the spin--orbit $W$ and for the tensor parameters. The reported values are in 
units of MeV fm$^5$.}
\end{table}

\begin{figure}[htb]
\begin{center}
\includegraphics[width=8cm]{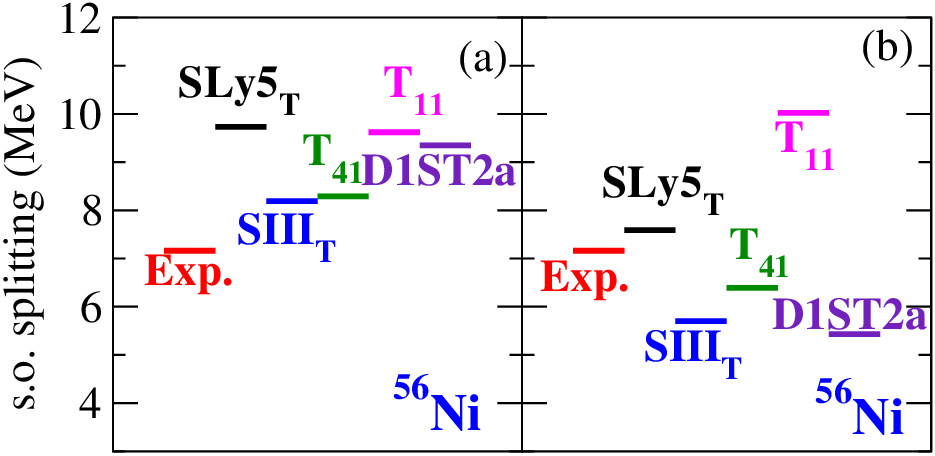}
\end{center}
\caption{(Color online) (a):  Spin--orbit splitting for the nucleus $^{56}$Ni by using the refitted parameters 
$\alpha_{\rm T}$ and $V_{{\rm T}1}+V_{{\rm T}2}$ 
with $\beta_{\rm T}$, $V_{{\rm T}2}$ = 0; (b): same as in (a) with $\beta_{\rm T}$, $V_{{\rm T}2} \ne 0$.}
\end{figure}

\begin{figure}[htb]
\begin{center}
\includegraphics[width=8cm]{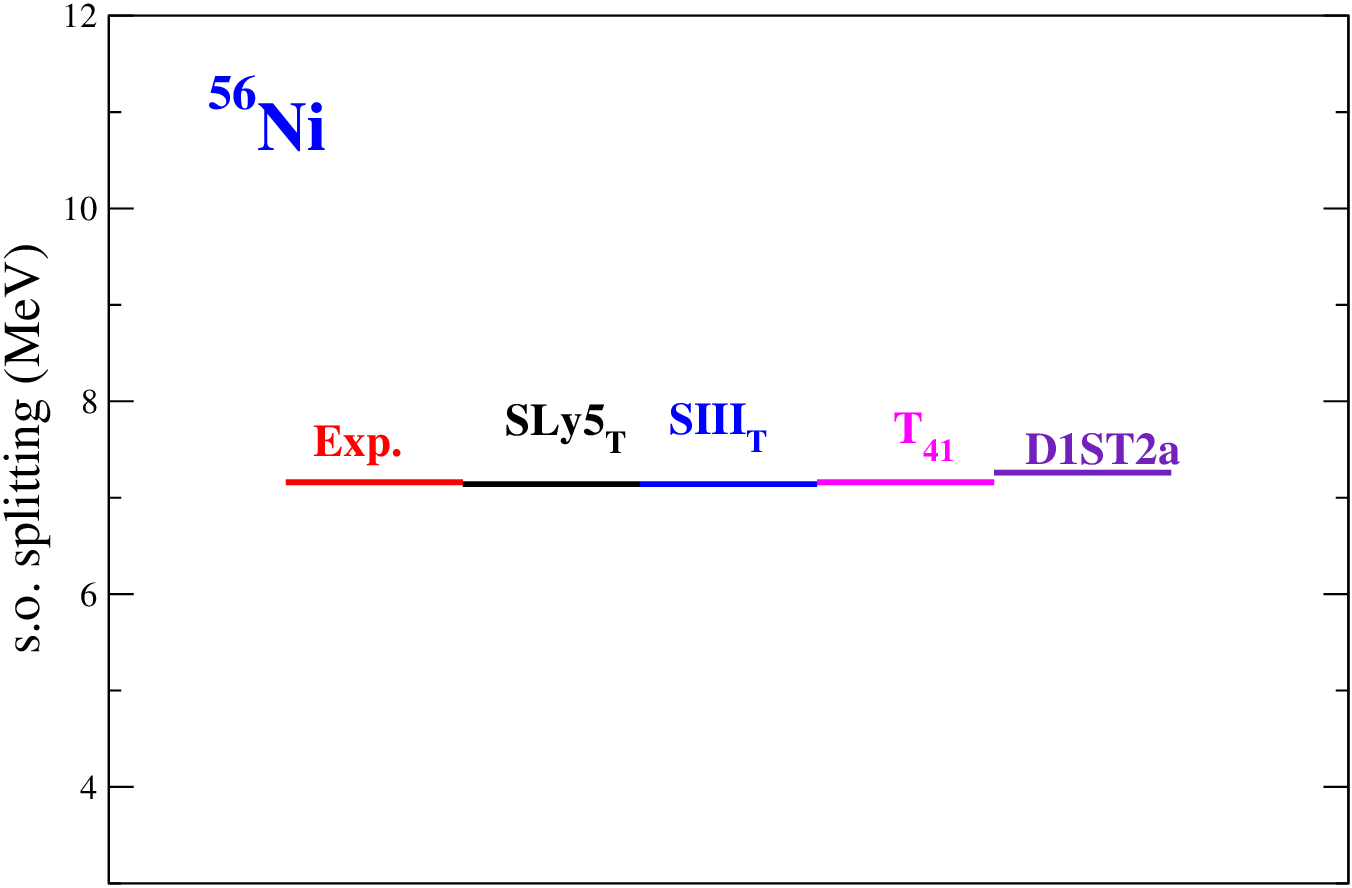}
\end{center}
\caption{ (Color online) Spin--orbit splitting for the nucleus $^{56}$Ni by using the reduced spin--orbit strength and the refitted parameters $\alpha_T$ and $V_{T1}+V_{T2}$, 
$\beta_{\rm T}$ and $V_{{\rm T}2}$.}
\end{figure}

To summarize the results reported in Table III, we observe that the spin--orbit parameter has to be systematically 
reduced for all the forces under study to provide the correct splitting in the nucleus $^{40}$Ca. The correct trend 
for the like--particle tensor effect is obtained only with those sets of parameters where $\alpha$ and 
$V_{{\rm T}1}+V_{{\rm T}2}$ are negative in the Skyrme and Gogny cases. The absolute values of these parameters have 
been modified in almost all cases (not for the SLy5$_{\rm T}$ case) with respect to their original values. The signs of the parameters are equal to those 
used in Ref. \cite{anguiano}, where the like--particle tensor behavior was analyzed in some applications.  As far as 
the neutron--proton effect is concerned, the value of $\beta$ has been in some cases increased and in other cases 
reduced; the value of $V_{{\rm T}2}$ has been decreased. The global result is that these parameters have to be 
positive. This sign provides the proton--neutron tensor mechanism as expected according to the shell--model calculations 
of Ref. \cite{otsu}. This was already underlined in Ref. \cite{moreno}. The opposite sign would lead to the opposite behavior. 

On the basis of these phenomenological adjustments and arguments, 
 the acceptable signs of the parameters can be identified and the regions of values can be indicated in a qualitative way. 
This implies first that 
several existing Skyrme parametrizations that include the tensor force should not be employed: in particular, 
all the parametrizations of  Ref. \cite{lesinski} where $\alpha$ is positive or $\beta$ is negative cannot reproduce 
the phenomenological results.  
For the Gogny case, the present work indicates that the parameters $V_{{\rm T}1}+V_{{\rm T}2}$ and $V_{{\rm T}2}$ should 
have the same sign as the corresponding Skyrme parameters $\alpha$ and $\beta$. 

We see from the new parameters shown in Table III that the 
strength that controls the like--particle effect of the tensor force ($\alpha_T$ or $V_{{\rm T}1}+V_{{\rm T}2}$) is always larger  than the value of the parameter responsible for the neutron--proton effect of the tensor force ($\beta_{\rm T}$ or $V_{{\rm T}2}$).
The ratio between the two strengths is equal to -1.4, -3.1, -3.9, and -1.2 for the new parametrizations SLy5$_{\rm T}$, SIII$_{\rm T}$, T$_{41}$, and D1ST2a, respectively. In the original cases, the four ratios were equal to -1.4, -1.5, -1.9, and -0.2, respectively. These ratios have been strongly modified for the cases SIII$_{\rm T}$, T$_{41}$ and D1ST2a. In particular, the most important change is in the case of the Gogny interaction, where the ratio was originally less than 1. With the present adjustment, this ratio becomes similar to that obtained with the interaction SLy5$_{\rm T}$.
This result is interesting and was not necessarily expected because the ranges of 
the tensor force are different in the Skyrme and Gogny models. This indication can also be useful for a future 
adjustment of a full Gogny parametrization including a pure tensor and a tensor--isospin term. 

As already stressed, these adjustments are qualitative for two reasons: (i) they are done at the simple HF level; (ii) the other parameters of the forces are not readjusted within a global fitting procedure. We have thus checked whether, even within a picture that is not quantitative, the masses predicted with the new sets of parameters are not dramatically shifted from the original values. 
We report in Table IV the binding energies obtained with the original and with the readjusted interactions for the three nuclei under study. Since in the original cases SIII$_{\rm T}$, SLy5$_{\rm T}$ and D1ST2a  of Refs. \cite{stancu2007}, \cite{colo}, and \cite{anguiano} the tensor part is not obtained with a global fitting procedure, but is adjusted on top of existing sets of parameters, we will show for these cases the binding energies obtained with the original parametrizations SLy5 \cite{chabanat}, SIII \cite{beiner} and D1S \cite{berger} that do not contain the tensor force. In the original SLy5 case, the  $J^2$ contributions generated by the velocity--dependent terms of the interaction are included in the Hamiltonian density. 

\begin{widetext}
\begin{center}
\begin {table} 
\begin{tabular}{ccccc}
\hline
                                                   & SLy5$_{\rm T}$ refitted  & SIII$_{\rm T}$ refitted & T$_{41}$ refitted & D1ST2a refitted \\
       & compared with SLy5 \cite{chabanat} & compared with SIII \cite{beiner} & compared with T$_{41}$ \cite{lesinski} &compared with D1S \cite{berger} \\
\hline 
\hline
Original $^{40}$Ca       & 344.07        & 341.88         & 339.79   &  342.09 \\  
Refitted $^{40}$Ca       & 343.83         & 341.62          & 339.44    & 344.75 \\
\hline
Original $^{48}$Ca     & 415.92       & 418.22        & 418.86 &  417.36 \\  
Refitted $^{48}$Ca &   415.91      & 415.55        & 411.07   & 414.58 \\
\hline
Original $^{56}$Ni              & 482.68      & 483.64       & 481.26   &  484.42 \\  
Refitted $^{56}$Ni             & 473.74       & 474.90        & 469.06   &   472.50 \\
\hline
\end{tabular}
\caption {Binding energies in MeV obtained with the different interactions for the nuclei $^{40}$Ca, $^{48}$Ca and $^{56}$Ni.}
\end{table}
\end{center}
\end{widetext}

We observe that the refitted sets induce modifications in the binding energies, as obviously expected. Anyway, these modifications (which are for instance important for the nucleus $^{56}$Ni) are not dramatic and one could expect that a future global fit of all the parameters would help us in reducing them. 
Of course, the objective of this global fit would not be to obtain again the same values of the binding energies found with the original forces, but to obtain reasonable values with an adjustment done on the experimental masses, without spoiling the good properties of the tensor part. 

Since the three nuclei under study are all medium--mass systems, we have also checked that the binding energies remain in an acceptable range of values also for the nuclei $^{16}$O, $^{90}$Zr and $^{132}$Sn, that are located in other regions of the nuclear chart. In the case of $^{16}$O, that is a spin--saturated nucleus, this check concerns only the modification of the spin--orbit part. 
The results are reported in Table V. One can draw the same conclusions as already done for $^{40}$Ca, $^{48}$Ca and $^{56}$Ni. 

\begin{widetext}
\begin{center}
\begin {table} 
\begin{tabular}{ccccc}
\hline
                                                   & SLy5$_{\rm T}$ refitted & SIII$_{\rm T}$ refitted  & T$_{41}$ refitted & D1ST2a refitted \\
       & compared with SLy5 \cite{chabanat} & compared with SIII \cite{beiner} & compared with T$_{41}$ \cite{lesinski} &compared with D1S \cite{berger} \\
\hline 
\hline
Original $^{16}$O       & 128.38        & 128.20         &  125.76 &  129.91 \\  
Refitted $^{16}$O       & 128.21        &  128.05       &   125.51  & 129.69 \\
\hline
Original $^{90}$Zr     & 783.30      & 782.65       &  786.85  &  786.55 \\  
Refitted $^{90}$Zr &    782.60   &  778.83     & 776.67  & 782.37 \\
\hline
Original $^{132}$Sn              & 1103.91      & 1105.85       &  1102.45  &  1105.15 \\  
Refitted $^{132}$Sn             &  1092.14    &   1093.66    & 1084.67   &   1088.85 \\
\hline
\end{tabular}
\caption {Binding energies in MeV obtained with the different interactions for the nuclei $^{16}$O, $^{90}$Zr and $^{132}$Sn.}
\end{table}
\end{center}
\end{widetext}
 
\section{Conclusions}

In this work we have used a fitting protocol introduced in Refs. \cite{zalewski2008,zalewski2009} starting from existing 
Skyrme and Gogny effective interactions that include a tensor force. First, the spin--orbit parameter of the chosen 
existing parametrizations (six Skyrme and one Gogny sets of parameters) is modified (reduced in all cases) to reproduce 
the neutron $1f$ spin--orbit splitting in the fully spin--saturated nucleus $^{40}$Ca. 
This case of $^{40}$Ca has been chosen as an example of isotope where the standard spin--orbit strength does not provide the correct splitting and the tensor force has no effects. This indicates the necessity of checking  the spin--orbit strength before performing the adjustment of the tensor parameters.

By employing this parameter, the like--particle tensor effect is tuned on the nucleus $^{48}$Ca to reproduce the 
neutron $1f$ spin--orbit splitting. This nucleus is spin--saturated for protons and the neutron--proton tensor effect is 
thus negligible. The parameter $\alpha$ ($V_{{\rm T}1}+V_{{\rm T}2}$) for the Skyrme (Gogny) case is readjusted for those 
interactions where these parameters are negative. On the other side, the Skyrme interactions where $\alpha$ is positive 
are excluded because they do not provide the correct behavior with respect to the experimental values. Finally, by keeping the new spin--orbit and 
like--particle tensor parameters, we tune the neutron--proton tensor contribution on the nucleus $^{56}$Ni that is 
spin--unsaturated. Only the Skyrme interactions where $\beta$ is positive are retained because $\beta \le 0$ leads to a 
wrong tensor effect. The parameters $\beta$ and $V_{{\rm T}2}$ are then modified to reproduce the experimental 
splitting. In some cases these parameters are reduced and in other cases increased with respect to their original values. 
The general result is that these parameters have to be positive. It is worth observing that the 
signs of the tensor Skyrme parameters are the same as the corresponding Gogny parameters, in spite of the fact that the range of the tensor force is different in the two cases. It is also interesting to
mention that, in the cases SLy5$_{\rm T}$ and D1ST2a, the ratio between the parameters that governs
the like--particle and the neutron--proton channels is very similar (-1.4 and -1.2, respectively). 
  
With the present phenomenological arguments, based on the fitting protocol of Refs. \cite{zalewski2008,zalewski2009}, all 
the Skyrme parametrizations of Ref. \cite{lesinski} where $\alpha$ is positive or $\beta$ is negative have to be excluded. 
The found regions of acceptable values for the parameters in the Gogny case can be a very useful starting point for a 
future full adjustment of the Gogny interaction in cases where a complete (pure tensor plus tensor--isospin) tensor force 
is included. The objective of this global fit would be to provide reasonable values for the binding energies by keeping reasonably good properties related to the tensor contribution.

%
%
%

\end{document}